# Pre-trained Transformer-models using chronic invasive electrophysiology for symptom decoding without patient-individual training

1st Timon Merk
Movement disorder and Neuromodulation Unit
Charité University Medicine Berlin
Berlin, Germany
timon.merk@charite.de

2nd Saeed Salehi
Machine Learning Group
Technical University Berlin
Berlin, Germany
salehinajafabadi@tu-berlin.de

3rd Richard M. Köhler
Movement disorder and Neuromodulation Unit
Charité University Medicine Berlin
Berlin, Germany
koehler.richard@charite.de

4th Qiming Cui
Department of Bioengineering and Therapeutic Sciences
University of California San Francisco
San Francisco, United States of America
Gavin.Cui@ucsf.edu

5th Maria Olaru
Department of Neurological Surgery
University of California San Francisco
San Francisco, United States of America
molaruna0@gmail.com

6th Amelia Hahn
Department of Neurology
University of California San Francisco
San Francisco, United States of America
amelia.hahn@ucsf.edu

7th Nicole R. Provenza
Department of Neurosurgery
Baylor College of Medicine, Houston
Houston, United States of America
Nicole.Provenza@bcm.edu

8th Simon Little
Department of Neurology
University of California San Francisco
San Francisco, United States of America
Simon.Little@ucsf.edu

9th Reza Abbasi-Asl
Department of Bioengineering and Therapeutic Sciences
University of California San Francisco
San Francisco, United States of America
Reza.AbbasiAsl@ucsf.edu

10th Phil A. Starr*
Department of Neurological Surgery
University of California San Francisco
San Francisco, United States of America
Philip.Starr@ucsf.edu

11th Wolf-Julian Neumann*
Movement disorder and Neuromodulation Unit
Charité University Medicine Berlin
Berlin, Germany
julian.neumann@charite.de
*shared authors

*Abstract*— Neural decoding of pathological and physiological states can enable patient-individualized closed-loop neuromodulation therapy. Recent advances in pre-trained large-scale foundation models offer the potential for generalized state estimation without patient-individual training. Here we present a foundation model, trained on chronic longitudinal deep brain stimulation recordings spanning over 24 days. Adhering to long time-scale symptom fluctuations, we highlight the extended context window of 30 minutes. We present an optimized pre-training loss function for neural electrophysiological data, that corrects for the frequency bias of common masked auto-encoder loss functions due to the 1/f power law. We show in a downstream task decoding of Parkinson's disease symptoms with leave-one-subject out cross-validation without patient-individual training.

## I. Introduction

Deep learning breakthroughs using the transformer architecture have enabled performance generalization with few -or zero-shot learning using massive amounts of data in different modalities, such as text or images [1]. First remarkable demonstrations of invasive and non-invasive neural electrophysiological time-series foundation models were presented with clinical applications, such as Alzheimer's disease, epilepsy or depression disease classification [2], [3]. Yuan et al presented a transformer foundation model using electroencephalographic (EEG) and stereotactic EEG (sEEG) recordings using in-clinic recordings. For many stationary and short-term recordings, these specific architecture parameters may be optimal. However, chronic long-term recordings streamed in naturalistic at-home settings from brain implants have brought new insights [4]. Currently, no transformer model has been developed using chronic, long-term DBS electrode recordings that can be used to improve clinical analysis. Even though pretrained models using invasive neural recordings have been developed previously, the underlying data was obtained exclusively from in-hospital recordings. The established invasive neuromodulation therapy deep brain stimulation (DBS), used in clinical routine to treat movement disorders such as Parkinson's disease (PD), offers a unique opportunity to longitudinally record invasive brain signals through implanted stimulation electrodes in the patients natural at-home environment. First chronic home-streamed recordings have recently become feasible through sensing enabled brain implants that enable longitudinal access into invasive electrophysiological brain signals across varying symptom and behavioural states [5]. Neural symptom decoding based on these recordings could enable closed-loop stimulation to treat each symptom precisely when it occurs [6]. Disentangling the superimposed physiological and pathological states remains a major challenge for accurate state identification within closed-loop neuromodulation. Physiological states resemble healthy or natural behaviour, such as movement, speech or sleep, while pathological states are disease specific, for example slowness of movement (bradykinesia), involuntary motor output (dyskinesia) or tremor in Parkinson's disease. A key advantage of chronic recordings is the extended temporal window they provide, enabling detection of long-term neural trends and their relationship to clinical symptoms. For example, intracranial recordings from DBS electrodes follow circadian rhythms [7], and the disruption of long-term neural periodicity could predict clinical response in psychiatric disorder patients [8]. Here, we used chronically streamed deep brain stimulation recordings from Parkinson's disease patients to pre-train a transformer-based foundation model. Due to the inherent temporal modulation of symptoms and brain signal features, we concatenated electrophysiological and temporal features within the token representation. As a result of the inherent 1/f power-law, present in neural recordings [9], pre-training based on common loss functions, such as mean absolute or mean squared error, biases the computed loss towards low frequencies. Thus, we present an adapted pre-training loss function including logarithmical scaling. We show that important symptoms of Parkinson's disease, such as bradykinesia or dyskinesia could be decoded by fine-tuning the pre-trained model within leave-one-subject-out cross-validation.

## II. METHODS

### A. Data acquisition

We enrolled 16 patients (2f) with Parkinson's disease that were implanted bilaterally with the Medtronic Summit RC+S investigational deep brain stimulator at the University of California San Francisco. We recorded local field potential recordings with a sampling rate of 250 Hz through the implanted stimulation electrodes from either the globus pallidus or subthalamic nucleus targets within the basal ganglia. Deep brain stimulation channels were bipolarly referenced to adjacent contacts. In addition to electrophysiological recordings, we obtained wearable bradykinesia and dyskinesia symptom estimates with a sampling interval of 2 min through an optimized smart watch for Parkinson's disease (PKG Health) [10]. We grouped data into consecutive 30 min sequences, resulting overall in 11935 sequences (mean subject sequences: 398 ± 240).

### B. Model architecture

In this work we used the transformer-encoder architecture to process neural recordings. In brief, a transformer processes sequences in parallel by combining token embeddings with positional information and using self-attention to capture relationships between all tokens, enabling it to model complex dependencies efficiently [11]. To represent each token, we computed the log10-transformed Welch spectrogram aligned to each 2 min wearable symptom window, and concatenated the "hour" of the day, which was represented as an ordinal variable ranging from 0 to 23. We joined 15 tokens as a continuous sequence (total duration: 30 min), appended a classification/register (CLS) token to the sequence and added learnable positional encoding [12] (Figure 1a). The linear projection layer maps token-features from the input dimension 125 into the embedding space with model dimension 64 through a linear layer. The learnable positional encoding, added to the input embeddings, allow the transformer to incorporate sequential order information. Thus, the linear projection layer serves both to reduce dimensionality and to embed contextual temporal information into each token. The internal building block of the transformer-encoder is then a repeated sequence of attention and feed-forward layers (Figure 1b). The attention mechanism assigns context-dependent weights to all elements in the sequence, enabling to capture long-range relationships irrespective of their position (Figure 1c). We parametrized the model architecture with a model dimension of 64, a feedforward dimension of 32, 4 attention heads and 2 transformer layers (total 51456 parameters). For pre-training we specified a learning rate of 1e-4 using the AdamW optimizer [13] with beta1 0.9 and beta2 0.95, 100 epochs training duration and a batch size of 50. We pre-trained the model using self-supervised learning as a masked autoencoder. For this purpose, we randomly set 30 % of all tokens to zero and computed the logarithmically scaled mean absolute error between the true masked neural tokens and linear fitted predictions from the transformer latent representations (Figure 2). We visualized the pre-trained transformer embeddings using t-SNE (scikit-learn implementation v1.6.1 [14], 2 components, 1000 iterations, 40 perplexity) and visualized the bradykinesia and dyskinesia symptom score representations.

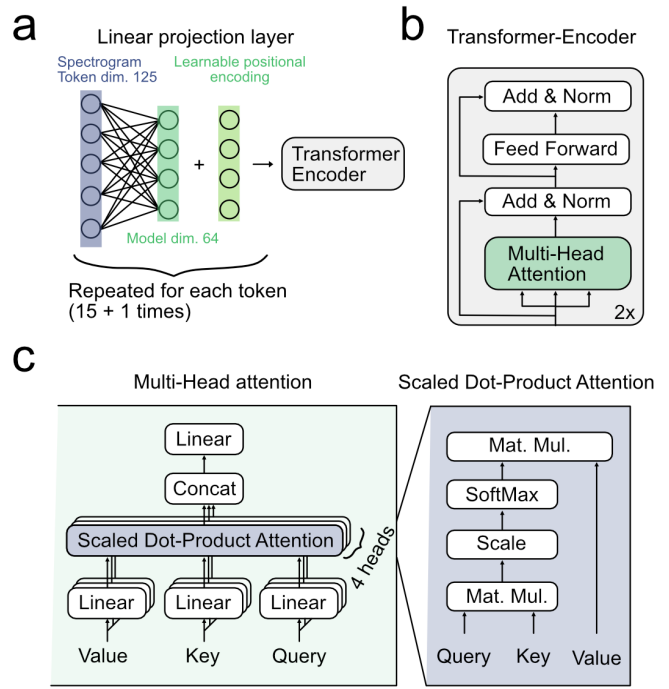

Fig. 1. Linear projection layer and transformer-encoder schematic. a) The linear projection layer reduces the spectrogram dimension from 125 to the model dimension 64 for each token. The learnable positional encoding is a token-wise vector addition. b) The internal transformer-encoder block schematic is displayed including c) multi-head attention to process information from multiple representation subspaces in parallel, enabling the model to capture contextual relationships within the sequence.

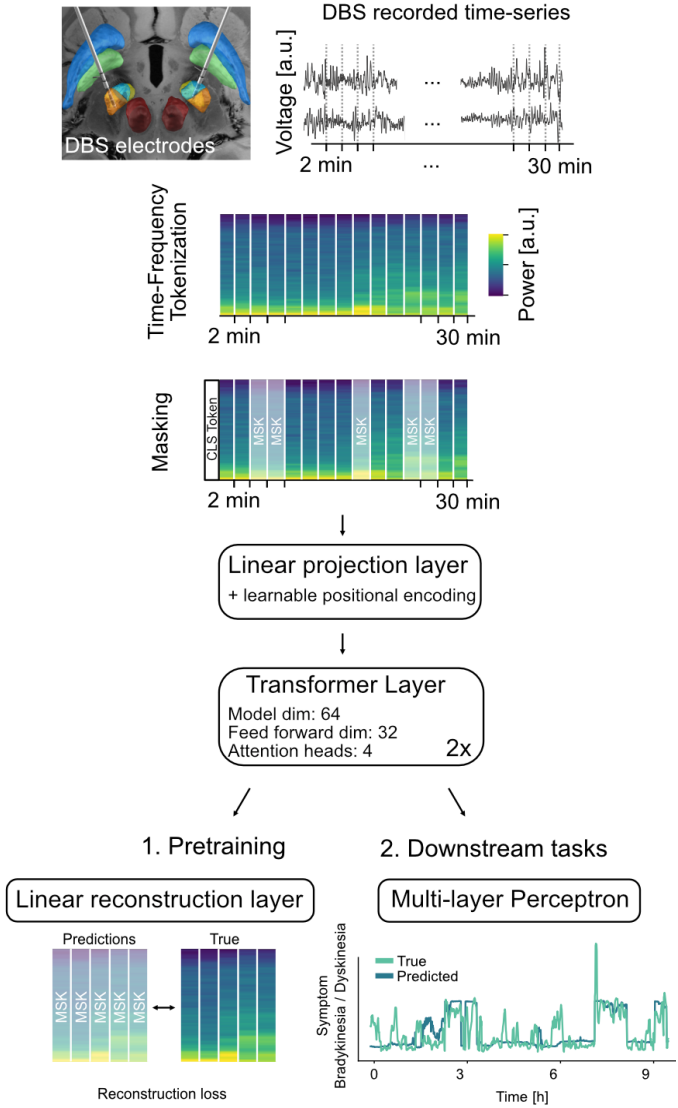

Fig. 2. Pre-training and downstream fine-tuning schematic for the transformer-based foundation model based on DBS local field potentials. From the time-series input, a spectrogram was computed for each two minute interval. Through a linear projection layer the token were fed to a transformer-encoder. The transformer output was then trained to predict masked spectrogram inputs during pre-training and wearable symptom estimates during fine-tuning in leave-one-subject out cross-validation.

### C. Loss-function scaling

Neural data is commonly characterized by periodic and aperiodic components [9]. The aperiodic component follows a 1/f distribution scaled by a factor $\beta$:

$$S(f) \propto \frac{1}{f^\beta} \qquad (1)$$

The power spectral density (PSD) is then represented as the sum of the aperiodic component and a set of periodic peaks, approximated as gaussian distributions [9]. The aperiodic 1/f component can have consequences for several applications. By training a PSD-based masked autoencoder, the commonly used mean absolute or mean squared error loss functions bias the loss frequency dependent towards low over high frequencies. The most common methods proposed for the correction of the 1/f spectral decay, rely on the separation of periodic and aperiodic components, such as specparam [9] and IRASA [15], by either iterative fitting or resampling, which would inflate computational cost for pre-training.

Hence, we approximated the aperiodic spectral component by a logarithmic function with base 10 and verified that the negative approximate log function roughly aligns with the mean log10 transformed power spectral density $P \in \mathbb{R}^{N \times F}$ across all spectra $N$ and Fourier coefficients $F$. We then defined the vector $\vec{p}$ as the average of $P$. The mean PSD and loss scaling function is shown in Figure 3. Thus, a constant could be approximated for each frequency bin $f_i$:

$$const \approx \log(f_i) - P(f_i)\ \forall i \qquad (2)$$

Next, we defined the scaling vector $\vec{k}$ based on the log-transformed frequency vector and the mean spectrum:

$$\vec{k} = \log(\vec{f}) + \frac{1}{F}\sum_{f=1}^{F} p(f) \qquad (3)$$

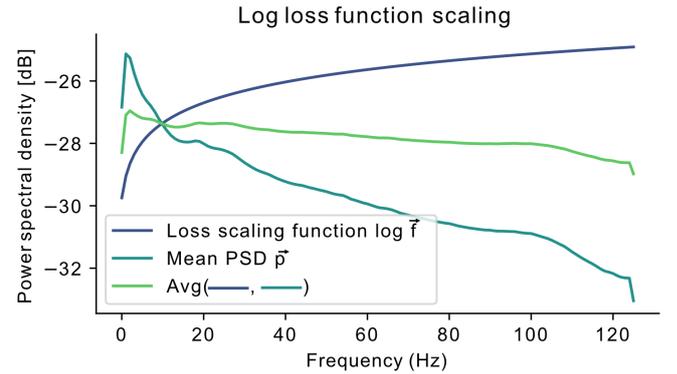

Fig. 3. Heuristic loss-function scaling to adjust for the 1/f neural power spectra distribution. The overall mean PSD was computed across patients. The log-transformed frequency vector approximates the inverse scaling of the mean PSD.

The vector $\vec{k}$ was then repeated to match the model dimension. We then defined the scaled mean absolute error masked autoencoder loss function given the model output $Y \in \mathbb{R}^{N \times F}$ and the predicted output $\hat{Y}$ as follows:

$$Loss = \frac{1}{N}\sum_{1}^{N}\left|KY - K\hat{Y}\right| \qquad (4)$$

We fine-tuned the base-model for each patient within a leave-one-patient out cross-validation using a two-layer multilayer perceptron (32 hidden units, ReLU activation function) for regression of symptom scores and specified early stopping at five epochs to prevent over-fitting.

### III. RESULTS

We pre-trained a transformer-based architecture as a masked autoencoder using large quantities of training data (24 days, 5 hours) in a self-supervised manner to predict masked tokens. Tokens were represented as 2 min power spectra within a sequence length of 30 min. We repeated pre-training within each cross-validation fold, excluding all data from the left-out subject. While the training error did not converge beyond 100 epochs, the validation error converged already at around 25 epochs for most patients (Figure 4). Using t-distributed Stochastic Neighbour Embedding (t-SNE), we verified that there were representational characteristics, that showed clear distinctions between symptoms (Figure 5). We speculate that this embedding space could represent further behavioural or pathological features that might be utilized for additional downstream tasks.

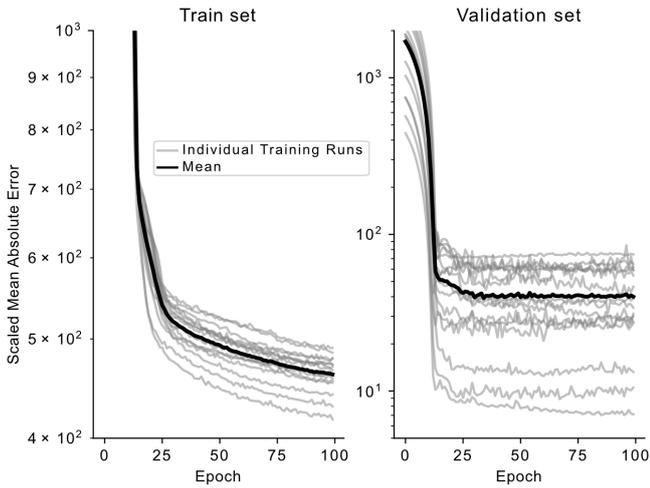

Fig. 4. Loss-functions for training and validation data. Light grey lines represent individual left-out patient losses, black lines mean across patients.

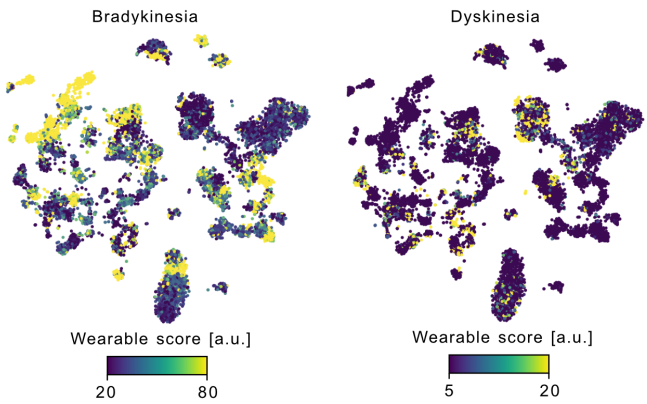

Fig. 5. Pre-trained t-SNE embeddings show clear distinctions between wearable bradykinesia and dyskinesia symptom states.

We then fine-tuned the foundation model by training a two-layer multilayer perceptron as a downstream architecture to predict wearable bradykinesia and dyskinesia symptoms (Figure 6a).

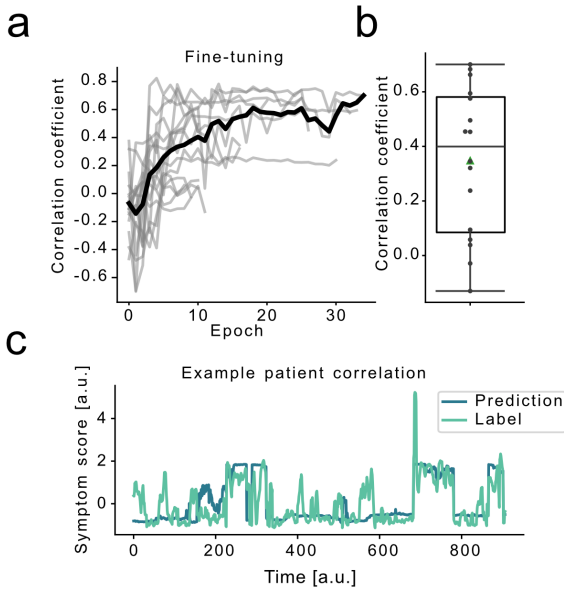

Fig. 6. a) Fine-tuning performance improvement with respect to training epochs. b) Overall leave-one-subject-out cross-validation Pearson correlation coefficient performances. c) Exemplar bradykinesia score predictions and true wearable labels.

Mean predictions resulted within the leave-one-subject-out cross-validation in Pearson correlation coefficients for bradykinesia of 0.35 ± 0.28 and for dyskinesia of 0.18 ± 0.17 (Figure 6b).

## IV. DISCUSSION

In this work we present a foundation model for deep brain stimulation local field potential recordings. We pre-trained the architecture as a self-supervised masked auto-encoder. Next to neural features, we used time of the day as an additional feature. This architectural component was motivated by several recent findings that showed clear oscillatory patterns related to circadian rhythms [7], [8]. Furthermore, we specified the segment length of a single token to be 120 seconds, and the overall sequence length to 30 minutes of neural recordings. This introduces variations compared to previously presented short-context window invasive neural foundation models [3]. For treatment of most neurological disorders, symptoms can fluctuate on long timescales. For example, in Parkinson's disease symptoms can exhibit fluctuations on slow drift across hours with medication and circadian cycles [7]. In theory, foundation models trained on diverse-enough training data could infer time-related information from neural characteristics. Nevertheless, the model architecture then still requires a sufficiently long context window in which slow fluctuations are represented. First approaches addressing this issue were presented for large language models, such as sparse attention or low-rank approximation [16], and could in future be applied to neural recordings to capture joint short -and long-term fluctuations. These long-range neural recording dependencies call for the collection of datasets that span days to weeks within the same patients to extract long-range rhythmicity that even extends to multi-day rhythms as shown in epilepsy [17]. Those patterns could also be directly estimated and represented within each token. For example, Provenza et al. reported that the predictability of neural periodicity tracts clinical response [8], which could be used as a feature in addition to oscillatory and temporal features. Multi-modal tokens were already successfully integrated in vision transformers [18], and could represent temporal, oscillatory, demographic, symptom-specific features for neural time-series transformers. In addition, wearable, and patient-individual reports could be included for training a joint embedding across modalities. Here, we already showed symptom decoding generalization across patients. However, by extending the dataset size and model complexity [19], different pathological and physiological modality components could potentially be extracted–these could guide the exploration and understanding of different neurological disorders, their underlying electrophysiological characteristics, corresponding biomarkers, and their interplay with healthy behavioural states. Zero-shot generalization across patients and diseases could then enable disease-agnostic features that could generalize across disease conditions and possibly re-define disease states and definitions.

## V. LIMITATIONS

While our study was based on one of the largest datasets of chronic DBS recordings to date, one limitation is the low sample size in comparison to the number of model parameters. Due to slow symptom fluctuation timescales, we adapted the sequence length and individual token segment length to be representative of typical Parkinsons's disease fluctuation time windows. Therefore, a single sample was represented by 30 min of consecutive recordings. This creates a need for the collection of larger datasets of chronic invasive recordings and symptom markers. Deep learning performances were shown to scale with dataset size, therefore the underlying data has to represent the diversity of neural and symptom fluctuations. Furthermore, the ground-truth symptoms were indirectly estimated in this study using wearable smartwatch recordings. While previous studies validated correlations with true clinician-rated symptoms [10], noise and movement artifacts might lead to biases. In addition, we present only downstream task predictions of two Parkinson's disease symptoms. There is a need to generalize recordings from multiple diseases and healthy states, extending the here presented applications.

## VI. Conclusion

Foundation models were recently demonstrated across different domains with outstanding performances with few - or zero-shot generalization capabilities. First invasive neural foundation models were previously trained using large datasets primarily obtained through short intra-operative recordings. Here, we present a foundation model trained on naturalistic at-home invasive deep brain stimulation recordings. We adapted previous neural time-series architectures by adding the hour of the day as an additional token representation, extended the context window to address long-range symptom fluctuations, and demonstrated within a fine-tuning task leave-one-subject-out cross-validation generalization of Parkinson's disease symptom decoding without patient-individual training.

## VII. Code Availability

All required code for pre-processing, pre-training and fine-tuning can be accessed through a GitHub repository: https://github.com/timonmerk/dbs_foundation_model.

## Acknowledgments

Investigational devices were provided at no charge by the manufacturer Medtronic™, but the manufacturer had no role in the conduct, analysis or interpretation of the study. Computation has been performed on the HPC for the Research cluster of the Berlin Institute of Health. The study was funded by Deutsche Forschungsgemeinschaft (DFG, German Research Foundation) – Project-ID 424778371 and a US-German Collaborative Research in Computational Neuroscience (CRCNS) grant from the German Federal Ministry for Research and Education and NIH (R01NS110424). WJN received funding from the European Union (ERC, ReinforceBG, project 101077060).